\documentclass[preprint,12pt,5p,authoryear]{elsarticle}

\usepackage{amssymb}
\usepackage{multirow}
\usepackage[colorlinks=true]{hyperref}\usepackage[normalem]{ulem} % for strikethroughs
\usepackage{xcolor}
\usepackage{amsmath}

\journal{Advances in Space Research}

\begin{document}

\begin{frontmatter}

\title{Characterizing the Radio Quiet Region Behind the Lunar Farside for Low Radio Frequency Experiments}

\author[CASA]{Neil Bassett\corref{cor1}}
\ead{Neil.Bassett@colorado.edu}
\cortext[cor1]{Corresponding author}

\author[CASA,Ames,USRA]{David Rapetti}

\author[CASA]{Jack O. Burns}

\author[CASA,Phys]{Keith Tauscher}

\author[Goddard]{Robert MacDowall}

\address[CASA]{Center for Astrophysics and Space Astronomy, Department of Astrophysical and Planetary Science, University of Colorado, Boulder, CO 80309, USA}
\address[Ames]{NASA Ames Research Center, Moffett Field, CA 94035, USA}
\address[USRA]{Research Institute for Advanced Computer Science, Universities Space Research Association, Mountain View, CA 94043, USA}
\address[Phys]{Department of Physics, University of Colorado, Boulder, CO 80309, USA}
\address[Goddard]{NASA Goddard Space Flight Center, Greenbelt, MD 20771, USA}

\begin{abstract}

Low radio frequency experiments performed on Earth are contaminated by both ionospheric effects and radio frequency interference (RFI) from Earth-based sources. The lunar farside provides a unique environment above the ionosphere where RFI is heavily attenuated by the presence of the Moon. We present electrodynamics simulations of the propagation of radio waves around and through the Moon in order to characterize the level of attenuation on the farside. The simulations are performed for a range of frequencies up to 100 kHz, assuming a spherical lunar shape with an average, constant density. Additionally, we investigate the role of the topography and density profile of the Moon in the propagation of radio waves and find only small effects on the intensity of RFI. Due to the computational demands of performing simulations at higher frequencies, we propose a model for extrapolating the width of the quiet region above 100 kHz that also takes into account height above the lunar surface as well as the intensity threshold chosen to define the quiet region. This model, which we make publicly available through a Python package, allows the size of the radio quiet region to be easily calculated both in orbit or on the surface, making it directly applicable for lunar satellites as well as surface missions.

\end{abstract}

\begin{keyword}
%% keywords here, in the form: keyword \sep keyword
Radio frequency interference \sep Low frequency \sep Earth-Moon system \sep Finite difference time domain
%% MSC codes here, in the form: \MSC code \sep code
%% or \MSC[2008] code \sep code (2000 is the default)
\end{keyword}

\end{frontmatter}

%%
%% Start line numbering here if you want
%%
% \linenumbers

%% main text
\section{Introduction}
\label{introduction}

Low radio frequency experiments have the potential to make scientific breakthroughs in a number of different fields in astronomy. One such application is to observe the 21-cm spectrum of neutral hydrogen (HI) in the early universe. 

%Around 400,000 years after the Big Bang the Universe cooled sufficiently for matter and radiation to decouple, allowing the first neutral atoms to form. This epoch of recombination can be observed through the Cosmic Microwave Background (CMB), at a redshift of z $\sim$ 1,100. Much later, at a redshift of z $\sim$ 10-15, the first stars and galaxies formed. Upcoming optical and infrared observatories such as the James Webb Space Telescope are expected to be able to observe these early generations of luminous objects. The period in between these two regimes, however, remains largely unobserved. This observational gap can be addressed by utilizing the 21-cm spectrum of neutral Hydrogen (HI), which was abundant in the early Universe after recombination.

\begin{figure*}
    \centering
    \includegraphics[width=0.8\linewidth]{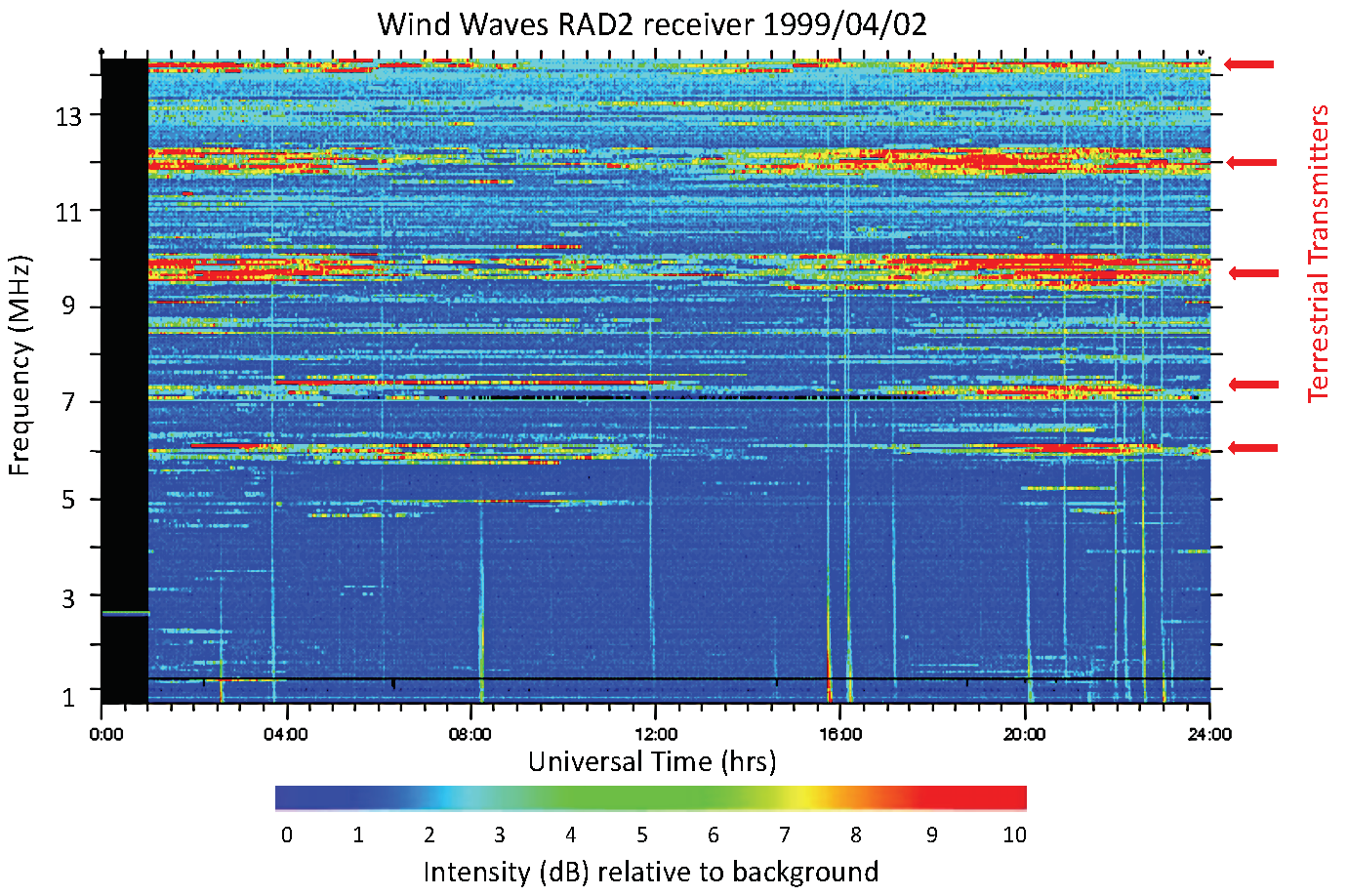}
    \caption{Data from the WAVES instruments on the Wind spacecraft taken from the vicinity of the Moon in 1999. The data show contamination from both terrestrial RFI (horizontal bands) and solar radio bursts (vertical lines). The intensity is measured relative to the galactic brightness. The intensity scale is limited to the maximum measured intensity, which is well below the saturation level of the instrument.}
    \label{fig:windwaves}
\end{figure*}

The hyperfine or ``spin-flip'' transition line of HI at a wavelength of 21-cm arises from the interaction between the magnetic moments of the electron and proton. 21-cm radiation from the early universe, which has a rest frequency of 1420 MHz, is highly redshifted such that the spectrum can be observed today between 10 and 200 MHz. The observed brightness temperature of the 21-cm transition is determined by both the temperature and neutral fraction of the hydrogen gas. Measurements of the 21-cm spectrum are therefore able to trace the history of the HI gas starting from the Dark Ages, after recombination when no luminous objects existed, through Cosmic Dawn, when gas clouds collapsed to form the first stars and galaxies, to the Epoch of Reionization (EoR), at which point the InterGalactic Medium (IGM) became fully ionized rendering the 21-cm line unobservable. 
For a more detailed overview of 21-cm cosmology, see \cite{Furlanetto:2006}, \cite{Pritchard:2012}, \cite{firstgalaxies}, and \cite{Morales:2010}.

Several ground-based experiments are current\-ly aiming to measure the 21-cm spectrum, with notable recent success. The Experiment to Detect the Global EoR Signature (EDGES) recently reported the detection of an absorption trough centered at 78 MHz (z $\sim$ 17), which constitutes the first ever observational measurement of the 21-cm spectrum \citep{Bowman:2018}. 21-cm spectrum measurements, in particular the Dark Ages portion of the spectrum at lower frequencies, may also be observed from space. The Dark Ages Polarimeter Pathfinder (DAPPER)\footnote{\url{https://www.colorado.edu/project/dark-ages-polarimeter-pathfinder/}} primarily aims to measure the 17-38 MHz portion of the 21-cm spectrum while in orbit behind the farside in order to test for physics beyond the standard cosmological model \citep{Burns:2019}.

%21-cm observations thus provide a method for studying the evolution of the early Universe during a time that is otherwise inaccessible. 21-cm experiments at these low radio frequencies can be performed by measuring either the power spectrum, which is determined by spatial anisotropies caused by the distribution of the HI gas, or the global signal, which is only the isotropic component. Power spectrum experiments rely on large arrays of interferometers to achieve the necessary angular resolution, while global signal experiments can be done with a single antenna. 
 
In addition to opening a new window to the early universe, low frequency observations may also play a key role in the search for habitable planets outside of the solar system. The presence of a planetary magnetosphere may play a significant role in retaining an atmosphere sufficient for habitability. The magnetic environment of the host star must also be taken into account. Coronal mass ejections and strong stellar winds may deplete the atmosphere of a planet that would otherwise lie in the habitable zone (\citealp{Kodachenko:2007}; \citealp{Lammer:2007}). This is particularly important for M dwar\-fs, which are very common and therefore may be good targets for exoplanet search programs, but are also known to be magnetically active compared to the Sun. Stellar magnetic events are often accompanied by radio bursts, which can be detected at low radio frequencies \citep{Cairns:2003}. Furthermore, planetary magnetic fields produce radio emission at low frequencies from cyclotron radiation due to auroral processes \citep{Zarka:1998}. Direct detection of this auroral radiation is the only method for direct observations of planetary magnetospheres. The Farside Array for Radio Science Investigations of the Dark ages and Exoplanets (FARSIDE) is a probe-class NASA-funded concept study to take advantage of the unique radio environment directly on the surface of the lunar farside. By deploying 128 antennas across a 10 km $\times$ 10 km area, FARSIDE will be able to image the entire sky between 200 kHz and 40 MHz \citep{farside}.
 
Ground based measurements at low radio frequencies are limited by both ionospheric effects and the presence of human generated radio frequency interference (RFI). Below $\sim$40 MHz the ionosphere distorts incoming radio signals (see e.g., \citealt{Vedantham:2014}; \citealt{Vedantham:2016}; \citealt{Rogers:2015}; \citealt{Datta:2016}). Below the plasma frequency, which depends on the electron density but is generally $\sim$9 MHz, the ionosphere is effectively opaque to electro\-magnetic radiation. Even above Earth's ionosphere, terrestrial RFI signals may still contaminate observations at low radio frequencies, as sho\-wn in observations taken by the WAVES instrument \citep{Bougeret:1995} in Figure \ref{fig:windwaves}. In addition, below $\sim$0.5 MHz the Earth produces intense auroral kilometric radiation (AKR) at heights up to 3 times the radius of Earth \citep{Gurnett:1974}. The farside of the Moon, however, offers a unique radio quiet environment where ionospheric effects are mitigated and contaminating radio emission is highly attenuated. Characterizing this radio quiet region through numerical simulations is therefore of great importance to low frequency radio experiments and is the basis of this work.

In order to determine the level of attenuation required to perform sensitive low frequency observations, we note that the maximum intensity of RFI relative to the galactic background, as shown in Figure \ref{fig:windwaves}, is $\sim$10 dB, which means that the intensity is a factor of 10 greater than the galaxy. The brightness temperature $T_{b}$ is defined as the temperature of a blackbody that would produce the equivalent amount of radiation observed. We also use $\nu$ to denote the frequency, $k$ for the Boltzmann constant, and $h$ for the Planck constant. In the Rayleigh-Jeans limit at low frequency (i.e. $h\nu \ll kT$), the power received by an antenna in a given frequency band $\Delta\nu$ is given by the following relation,
\begin{equation}
    P_{\nu} = kT_{b}\Delta\nu\,.
\end{equation}
Since intensity is simply the power transferred per unit area, the intensity within the frequency band is directly proportional to the brightness temperature,
\begin{equation}
    I_{\nu} \propto P_{\nu} \propto T_{b}\,.
\end{equation}
The brightness temperature of the RFI relative to the brightness temperature of the galaxy is thus also a factor of 10.

%This intensity can be converted to a brightness temperature through the blackbody spectrum as defined by the Planck function.
%\begin{equation}
%    I_{\nu} = \frac{2h\nu^3}{c^2}\frac{1}{e^{\frac{h\nu}{kT}} - 1}
%\end{equation}
%Thus, the brightness temperature can be calculated by the following equation:
%\begin{equation}
%    T_{b} = \frac{h\nu}{k\ln{\big(\frac{2h\nu^3}{I_{\nu}c^2} + 1\big)}} \approx \frac{I_{\nu}c^2}{2k\nu^2}
%\end{equation}
%\textcolor{green}{where the approximation holds at low frequencies (i.e. $\nu\ll \sqrt[3]{\frac{I_\nu c^2}{2h}}$)}. The brightness temperature of the RFI relative to the galaxy at a given frequency is then simply
%\begin{equation}
%    \frac{T_{RFI}}{T_{gal}} = \frac{I_{RFI}}{I_{gal}} = 10
%\end{equation}
Assuming that the galaxy has a temperature of 5000 K at 50 MHz and a spectral index of 2.5 \citep{Bowman:2018}, the temperature of the galaxy at 17 MHz\footnote{17 MHz is chosen as the reference frequency because it defines the low end of the proposed frequency band for DAPPER. The 15 mK RFI brightness temperature requirement is also driven by the DAPPER experiment.} is approximately 75,000 K. The temperature of the RFI is then 750,000 K. Assuming that we need the RFI brightness temperature to be $\leq$ 15 mK, then the amount of attenuation required is
\begin{equation}
\label{dB_thresh}
    10\log{\Big(\frac{750,000\ \textup{K}}{0.015\ \textup{K}}\Big)} = 77.0 \textup{ dB}\,.
\end{equation}
For the remainder of the paper we adopt -80 dB as the relative intensity threshold for the radio quiet region.

In previous work on the radio quiet environment of the lunar farside, \cite{Pluchino:2007} adopted the approximation that the Moon can be modeled as a straight edge such that the diffraction of radio waves around the edge can be calculated analytically. This model does not consider the electrical properties of the lunar material and may also be limited by the geometry of the straight edge, which does not take into account the presence of lunar material beyond the edge where the diffraction occurs. \cite{Takahashi:2003} took a similar approach to that presented below in this paper, utilizing a Finite Difference Time Domain (FDTD) algorithm to numerically simulate the electromagnetic environment. In this work, we utilize advances in software and computing resources to extend Takahashi's method by directly simulating higher frequency waves and including lunar properties such as topography and density profile using the same underlying simulation framework.
 
We organize this paper as follows. Section \ref{methodology} describes the numerical methods used to simulate the low frequency radio environment. Sections \ref{frequency}, \ref{topography}, and \ref{density} discuss the effect of frequency, topography, and density of the Moon on the results of the simulations, respectively, while in Section \ref{extrapolation} we model the results such that they can be extended to higher frequencies. Section \ref{calc_width} presents a Python package, which allows for direct calculations of the size of the quiet region based on the results of our simulations. Section \ref{Discussion} considers applications of our results for low frequency experiments and other lunar missions. Finally, we conclude by presenting a summary of the work in Section \ref{conclusions}.

\section{Methodology}
\label{methodology}

\subsection{Finite Difference Time Domain}
\label{fdtd}

The FDTD algorithm is a method for computational electrodynamics simulations, which discretizes both time and space onto a finite grid. A central-difference approximation to the partial derivatives, as outlined in \cite{Yee:1966}, is utilized to compute the time-evolution of Maxwell's equations. In order to perform the necessary discretization with second-order accuracy, FDTD methods calculate and sto\-re the E and B components at different locations on the grid, creating what is known as a Yee lattice. The fields are then be interpolated to a shared location before the output is stored or some value of interest such as energy density is calculated. As the size of the grid spacing and the time steps is decreased, the approximation approaches the continuous equations.

The MIT Electromagnetic Equation Propagation (Meep) software\footnote{\url{https://github.com/NanoComp/meep}} used for this project is an open-source implementation of the FDTD algorithm available in Python, Scheme, and C++ \citep{Oskooi:2010}. Meep includes a number of useful features such as dispersive materials, perfectly matched layers (PML) for simulating open boundary conditions, a variety of options for computing and outputting fields, and support for distributed memory parallelization.

The FDTD algorithm is only one of a wide variety of numerical methods for simulating electrodynamics. However, in contrast to other methods, which are often limited in their application, FDTD has the advantage of both versatility and simplicity. A wide variety of materials and conditions including nonlinear media can be simulated within the basic FDTD framework without altering the underlying algorithm. The algorithm also has the added benefit that it can be easily parallelized to take advantage of supercomputers such as NASA's Pleiades supercomputer, which was used for this project. For these reasons, the FDTD algorithm was determined to be the optimal method for performing simulations of the lunar environment at low radio frequencies.

\subsection{Geometry and Setup}
\label{geometry}

The rotational symmetry of the Earth-Moon system allows the simulation to be confined to two dimensions, significantly reducing the computational resources required. However, the large distance between the Earth and the Moon still poses a challenge due to the unrealistically high number of points needed to create a grid covering the entire system. Instead, we limit the grid to a 4000 $\times$ 4000 km region centered around the Moon. Reflection symmetry about the Earth-Moon axis allows us to reduce the number of grid points by half, further decreasing the necessary number of computations.

In order to approximate the electrical properties of the Moon, expressions for the electrical permittivity ($\varepsilon$) and the loss tangent ($\tan{\delta}$) at the lunar surface provided in \cite{sourcebook} were used. Both expressions depend on the density $\rho$ as follows:
\begin{equation}
\label{epsilon_equation}
    \varepsilon = 1.919^{\rho},
\end{equation}
\begin{equation}
\label{tand_equation}
    \tan{\delta} = 10^{(0.44\rho - 2.943)}\,.
\end{equation}
A useful quantity to calculate based on the electrical properties of the lunar material is the skin depth $\Delta$, which is defined as the depth at which the amplitude of the fields are reduced by a factor of $e$. For $\tan{\delta} \ll 1$,
\begin{equation}
\label{skin_depth_equation}
    \Delta \approx \frac{\lambda}{\pi\sqrt{\varepsilon}\tan{\delta}},
\end{equation}
where $\lambda$ is the vacuum wavelength. We used the average lunar density ($\bar{\rho}$ = 3.34 g cm$^{-3}$) to determine the average values for these electrical properties and then employed them in our primary simulations. For more details concerning the lunar density profile and its effect on the RFI quiet region, see Section \ref{density}.

Due to the two dimensional nature of the simulation, the orientation of the E and B fields must also be taken into account. In the Transverse Electric (TE) mode, the electric field lies within the plane of the simulation while the magnetic field is perpendicular to the plane. In the Transverse Magnetic (TM) mode the reverse occurs (i.e. the electric field is out of the plane). These modes become important when the waves interact with the dielectric material (i.e. the Moon). Simulating the Moon as a circle in a 2D simulation is analogous to simulating an infinite cylinder in 3D space, rather than a sphere. For the TM mode, the E field is oriented along the infinite axis of the cylinder. For a physical system in this orientation, charges in the cylinder would be excited by the E field and move along the infinite axis. For the TE mode, the electric field is tangential to the circle in the 2D plane so charges would move along the circumference. Since the latter scenario is a closer approximation to a 3D sphere, which has no infinite axis, we choose to use the TE mode in our simulations.

The edges of the grid need to mimic the boundary condition of the theoretically infinite open space that lies beyond the finite domain of the simulation. For this, we place a perfectly matched layer (PML), as introduced in \cite{Berenger:1994}, along the edge of the grid to absorb outgoing waves. Commonly used in many FDTD simulations, PMLs are constructed such that waves incident on the PML are prevented from reflecting off of the boundary back into the interior. 

Due to the small size of the domain of the simulation, the source creating the electromagnetic radiation must be placed close to the Moon. To simplify the simulation, the radiation is assumed to be in the form of a plane wave traveling horizontally in our simulations.  In general, three criteria must be met for the ``far field'' approximation, in which the wavefront appears locally as a plane wave, to hold:
\begin{equation}
    d \gg D, \quad d \gg \lambda, \quad d > \frac{2D^2}{\lambda},
\end{equation}
where $d$ is the distance between source and receiver, $D$ is the scale of the ``antenna,'' and $\lambda$ is the wavelength. Since there is no true receiver, we use the skin depth (as defined in Equation \ref{skin_depth_equation}) to estimate $D$, in which case we always satisfy the far field criteria. Note though that this plane wave approximation ignores the angular size of the Earth, an effect that may become more relevant for AKR emission, which emits up to several Earth radii. Note also, however, that when a spherical wavefront reaches the Moon, some of the radiation will be traveling at an angle opposing the direction in which diffraction is acting. The result is that the wave must be diffracted slightly more in order to reach the same point behind the lunar farside compared to a plane wave.

\begin{figure}
    \centering
    \includegraphics[width=\linewidth]{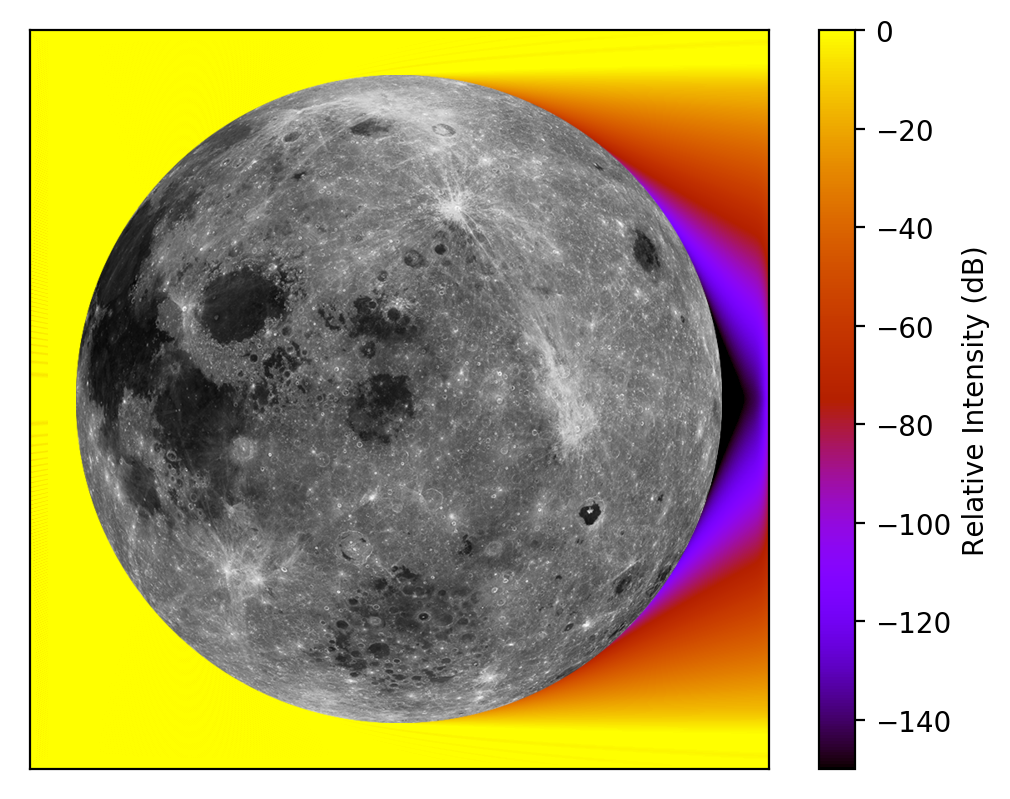}
    \caption{Results of a 4000 $\times$ 4000 km FDTD simulation at 30 kHz. RFI incident from the left is attenuated behind the Moon on the right. The intensity is calculated by comparing the simulation including the Moon to a simulation run without the Moon. An image of the Moon from the LROC WAC 643 nm reflectance mosaic \citep{Robinson:2010} is overlaid for illustrative purposes.}
    \label{fig:2d_sim}
\end{figure}

\begin{figure}[ht]
    \includegraphics[width=\linewidth]{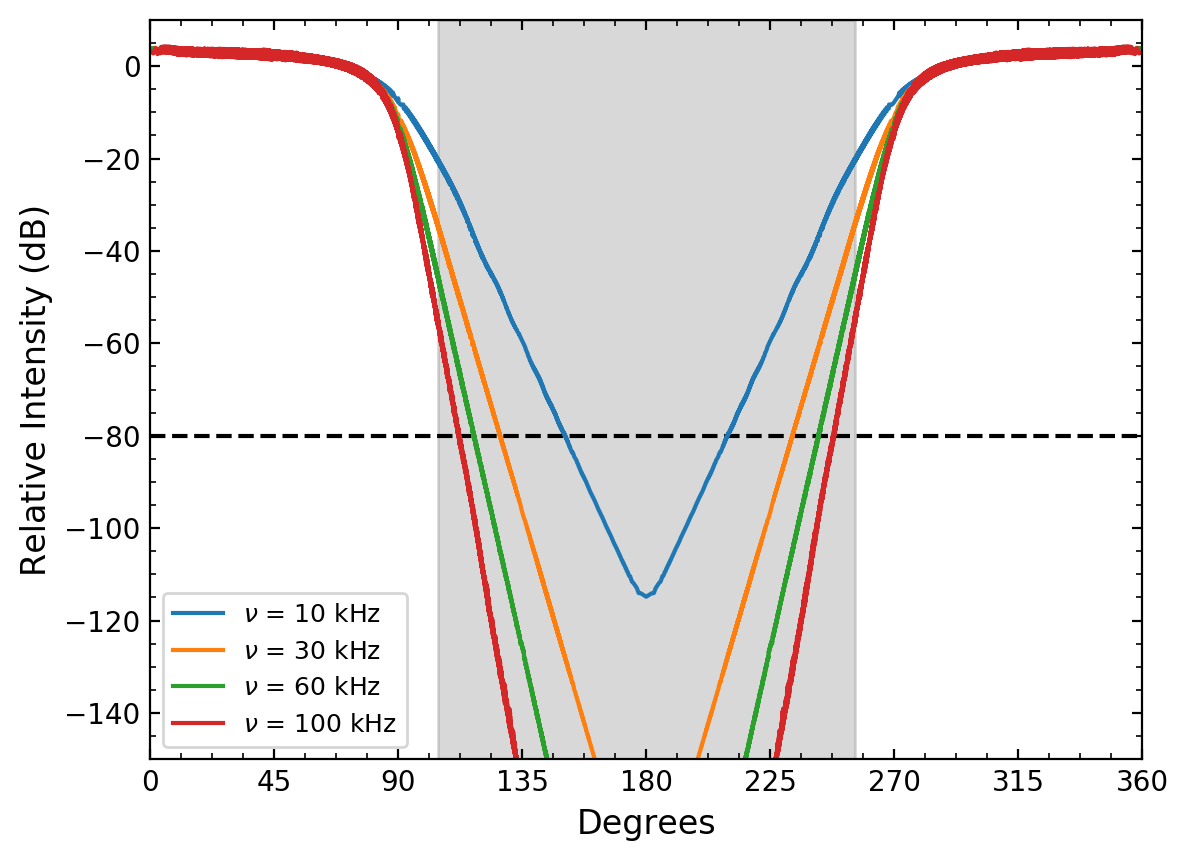}
    \caption{Relative intensity of radio frequency interference on the surface of the Moon for a range of frequencies. The geometric quiet region, neglecting the effects of diffraction, is indicated by the grey background between 90 and 270 degrees. The black dashed line designates the 80 dB suppression level, which we have adopted as the threshold for the quiet region (see Equation \ref{dB_thresh}). Even at frequencies as low as 10-100 kHz, there exists a large region on the farside where radio waves are sufficiently attenuated for sensitive low frequency experiments. Note that the lower bound of the y axis has been set to -150 dB due to possible numerical artifacts at intensities below this level.}
    \label{fig:RFI_surf}
\end{figure}

Although FDTD algorithms propagate the E and B fields through the simulations, characterizing the radio environment requires the intensity of the electromagnetic radiation at any given point. The intensity can be easily calculated from the fields after they are interpolated onto a common grid. In order to produce a meaningful intensity, the simulation is run twice, first without the Moon and then with the Moon included in the domain. In order to ensure that the intensity is not affected by the phase of the wave, the intensity is averaged over one full period. The relative intensity between the simulation with the Moon to the simulation without is able to directly measure the amount of attenuation provided by the Moon. It is important to remember then, that these results reflect only the effect of the Moon and not other factors such as free space path loss or attenuation by the Earth's ionosphere. The results of one such FDTD simulation are shown in Figure \ref{fig:2d_sim}.

\section{Results}
\label{results}

\subsection{Frequency Dependence}
\label{frequency}

Diffraction is a wave phenomenon, and as such it is dependent on the frequency of the radiation. This dependence is illustrated in Figure \ref{fig:RFI_surf} for a range of frequencies between 10 and 100 kHz. The amount of attenuation at a given location on the farside increases as the frequency of the radio waves is increased, following the fundamental expectation that higher frequency waves will diffract less. Due to computational constraints, performing FDTD simulations at higher frequencies, such as the 10-200 MHz region where the 21-cm spectrum resides, is not currently feasible. The resolution of the grid is chosen such that there are 8 pixels per wavelength within the dielectric material of the Moon in each simulation.
\begin{figure}[ht]
    \centering
    \includegraphics[width=\linewidth]{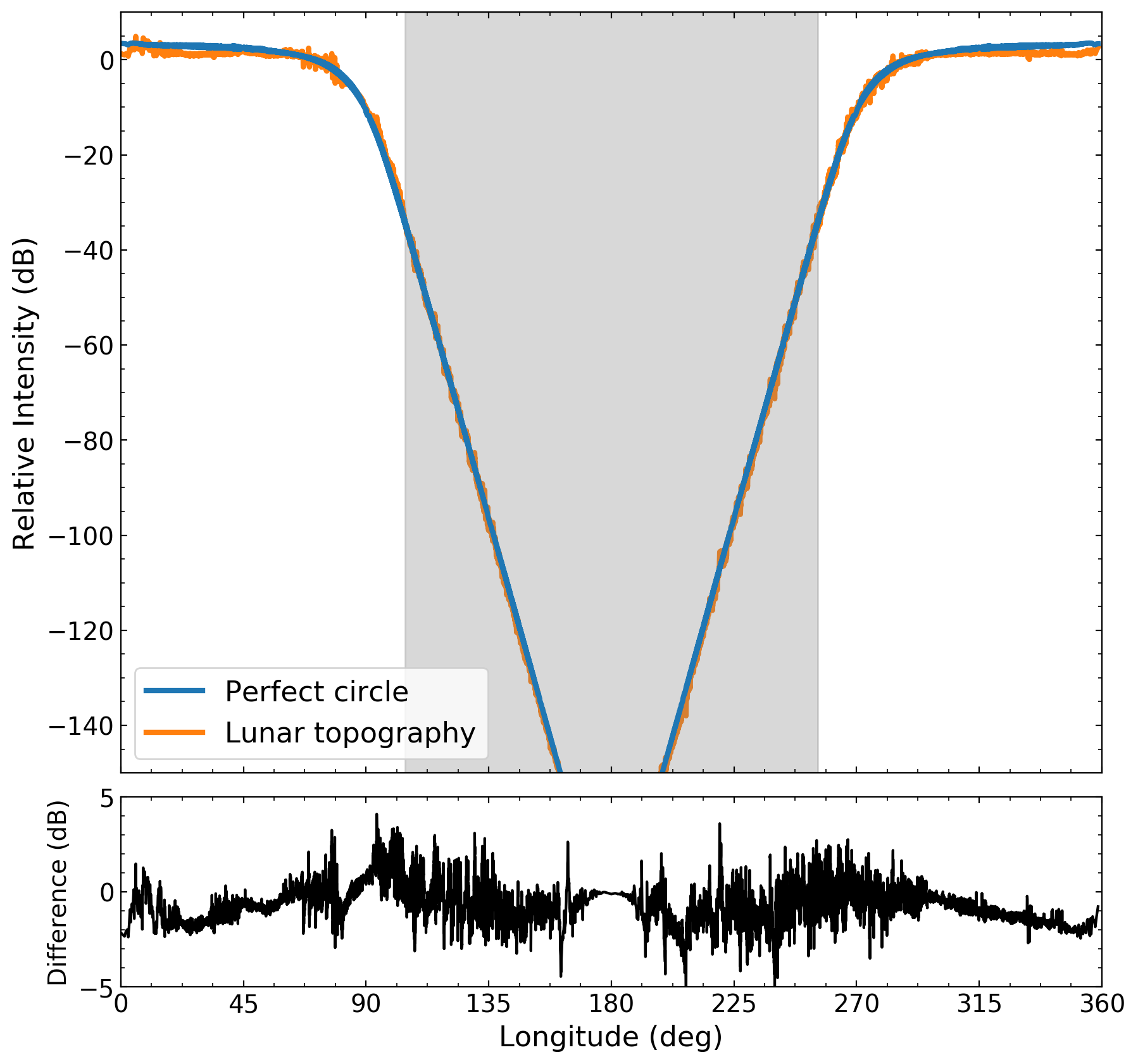}
    \caption{Comparison of two different FDTD simulations at 30 kHz, one in which the lunar surface is assumed to be smooth (blue) and one in which elevation data from the LOLA instrument was used to model the topography lunar surface (orange). As in Figure \ref{fig:RFI_surf}, the grey region indicates the geometric quiet region without diffraction. The results show that craters and mountains on the Moon create both constructive and destructive interference leading to differences in the intensity of the RFI. These small variations are on the order of a few dB.}
    \label{fig:RFI_craters}
\end{figure}
This choice of resolution also ensures that the grid spacing is much smaller than the skin depth $\Delta$, as defined in Equation \ref{skin_depth_equation}. As the frequency is increased, the wavelength becomes smaller and the distance between grid points must also be decreased. This requires more grid points to fill the domain, which in turn increases the computational load. The result is that simulating frequencies above 100 kHz is extremely expensive, even with supercomputing resources. However, lower frequency simulations are able to place a lower limit on the size of the radio quiet region, where RFI is attenuated by at least 80 dB, due to the fact that diffraction will be a smaller effect at higher frequencies. Note that although AKR is likely to dominate terrestrial RFI at these frequencies, since we will attempt to extrapolate these results to higher frequencies in Section \ref{extrapolation} we do not attempt to model any AKR-specific effects such as the increased source region or angular beaming.

\subsection{Topography}
\label{topography}

The results in Section \ref{frequency} assume that the Moon is a perfect sphere, which translates to a perfect circle in the two dimensional space of our simulations. In order to investigate the effect of the topography of the Moon on the amount of attenuation at radio frequencies, elevation data at the lunar equator from the Lunar Orbiter Laser Altimeter (LOLA; \citealt{Smith:2010}) instrument on the Lunar Reconnaissance Orbiter (LRO; \citealt{Chin:2007}) was used in an FDTD simulation to approximate the general topography of the Moon.

Figure \ref{fig:RFI_craters} shows the results of the simulation including the surface elevation data compared to the simulation assuming a perfect circle. Althou\-gh the topography of the Moon will certainly affect the propagation of RFI signals, the difference in intensity is small and for the most part can be ignored. This effect may play a bigger role at specific locations on the lunar surface such as inside very deep craters.

\subsection{Density Profiles}
\label{density}

\begin{table}
\centering
    \begin{tabular}[width=\linewidth]{llll}
    \multicolumn{4}{c}{Quiet Cone Widths ($\nu = 10$ kHz, -80 dB)} \\
    \hline\hline
     & 0 km & 50 km & 100 km \\%& 150 km \\
    \hline
    {\small Constant $\rho$ (blue)} & 58.7$^{\circ}$ & 42.5$^{\circ}$ & 32.4$^{\circ}$ \\% & 25.0$^{\circ}$ \\
    {\small Stepped $\rho$ (green)} & 62.2$^{\circ}$ & 45.8$^{\circ}$ & 35.2$^{\circ}$ \\%& 27.7$^{\circ}$ \\
    {\small Continuous $\rho$ (orange)} & 66.7$^{\circ}$ & 48.7$^{\circ}$ & 38.3$^{\circ}$ \\%& 30.9$^{\circ}$ \\
    \hline
    \end{tabular}
    \caption{Comparison of the width of the quiet region (at 80 dB attenuation) for a range of heights above the lunar surface. The colors refer to the curves in Figure \ref{fig:density_profiles_surface}. The data illustrate that using the constant density profile produces a lower limit on the size of the quiet region.}
    \label{density_table}
\end{table}

The electrical properties of the Moon depend on the density of the material, as shown in Equations \ref{epsilon_equation} and \ref{tand_equation}. The propagation and diffraction of radio waves behind the farside should then depend on the density profile of the lunar interior. The results shown from previous simulations assumed a constant density profile equal to the average density of the Moon ($\rho = 3.34$ g cm$^{-3}$). This constant profile significantly overestimates the density of the regolith near the surface and may underestimate the density deep in the interior.

\begin{figure}
    \centering
    \includegraphics[width=\linewidth]{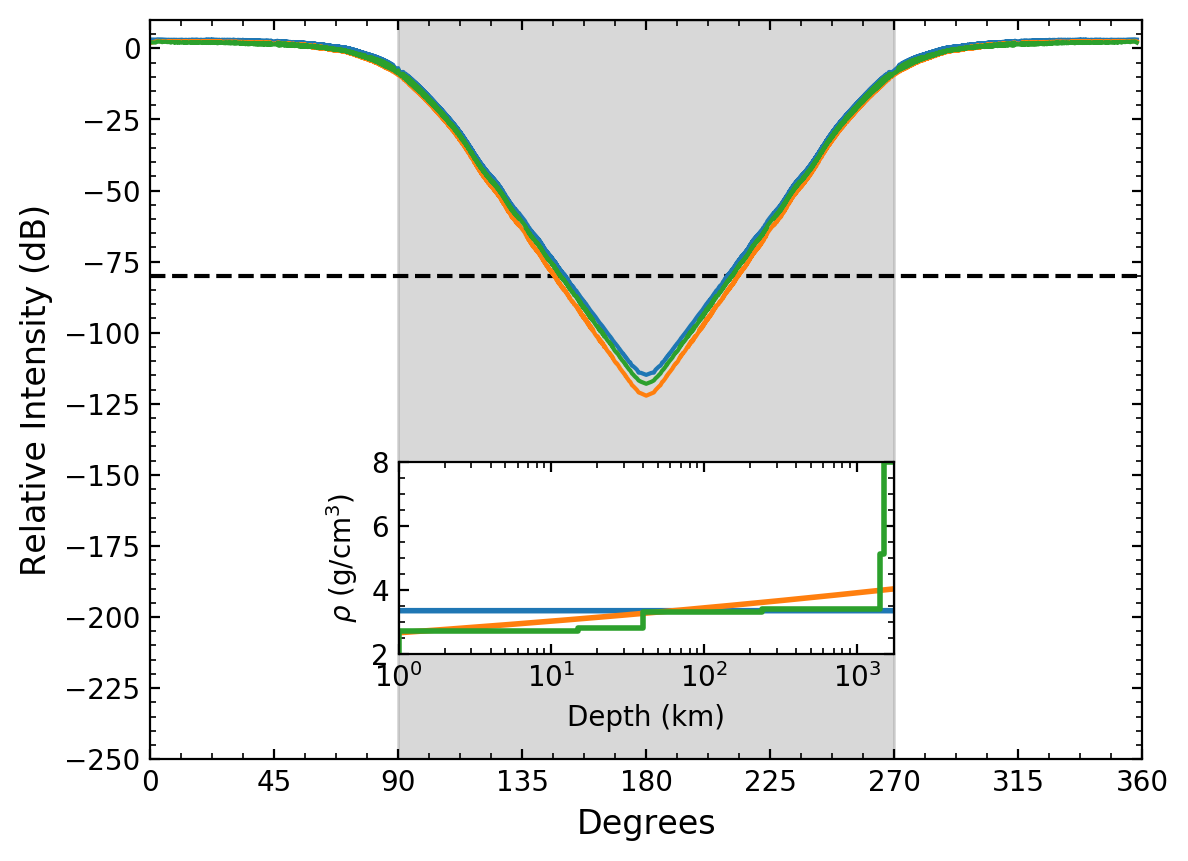}
    \caption{Comparison of three different density profiles for the interior of the Moon and their effect on the size of the radio quiet region at 10 kHz on the lunar farside. \textit{Inset:} Density as a function of depth for the constant average profile (blue), the continuous profile of \cite{sourcebook} (orange), and the stepped profile of \cite{Weber:2011} (green).}
    \label{fig:density_profiles_surface}
\end{figure}

As an initial test, we performed simulations with a constant density profile that instead of assuming the average density uses a much lower value representative of the lunar regolith. The consequence of using such a small value for the density is that the material has a very low loss and low frequency waves are largely able to pass through the material unattenuated. Assuming the electrical properties at the surface throughout the interior is clearly not a good approximation for the true electrical properties of the Moon.

\begin{table}
    \centering
    \begin{tabular}{llll}
        \multicolumn{4}{c}{Extrapolated Quiet Cone Widths (-80 dB)}\\
        \hline\hline
         &  surface & 50 km & 100 km \\%& 150 km \\
        \hline
        %10 kHz & 59.7$^{\circ}$ & 43.2$^{\circ}$ & 32.2$^{\circ}$ & 24.4$^{\circ}$ \\
        $\nu$ = 100 kHz & 135.7$^{\circ}$ & 111.6$^{\circ}$ & 101.6$^{\circ}$ \\%& 93.9$^{\circ}$ \\
        $\nu$ = 1 MHz & 162.6$^{\circ}$ & 136.8$^{\circ}$ & 126.6$^{\circ}$ \\%& 118.8$^{\circ}$ \\
        $\nu$ = 10 MHz & 172.1$^{\circ}$ & 146.0$^{\circ}$ & 135.6$^{\circ}$ \\%& 127.7$^{\circ}$ \\
        $\nu$ = 100 MHz & 175.4$^{\circ}$ & 149.4$^{\circ}$ & 138.8$^{\circ}$ \\%& 130.8$^{\circ}$ \\
        $\nu \rightarrow \infty$ & 180.0$^{\circ}$ & 151.4$^{\circ}$ & 140.6$^{\circ}$ \\%& 132.6$^{\circ}$ \\
        \hline
    \end{tabular}
    \caption{The width of the radio quiet region calculated for a range of heights and frequencies. -80 dB was used as the intensity threshold for all of the values in this table. The width in the limit that $\nu$ goes to infinity is equivalent to the width of the geometric cone (i.e. the dashed lines in Figure \ref{fig:model}).}
    \label{tab:width_table}
\end{table}

To investigate further we introduce a spatially variable density profile. Due to the difficulty of measuring density deep in the lunar interior, we compare two separate profiles. The first profile is a continuously varying function given by \cite{sourcebook}

\begin{equation}
    \rho = 1.39(z/\textup{cm})^{0.056}\ \textup{ g cm}^{-3},
\end{equation}

\noindent but is only based on measurements down to a depth of a few meters. The second profile is a stepped function given by \cite{Weber:2011} based on the analysis of lunar seismic data. The density profiles and their effect on the quiet region are compared in Figure \ref{fig:density_profiles_surface} and Table \ref{density_table}. The difference in the size of the quiet region for the three profiles is small, but it is important to note that the constant profile (blue) produces the least amount of attenuation and can be used as the `worst case' lower limit on the size of the quiet region. Although one might expect that the lower density (and thus lower loss) material near the surface in the orange and green profiles would produce a smaller quiet region, the higher density material deeper in the interior has a greater loss, which counteracts the surface material and increases the size of the quiet region. For this reason, we choose to report simulation results for the constant density profile in the remainder of the paper.

\begin{figure}[h]
    \centering
    \includegraphics[width=\linewidth]{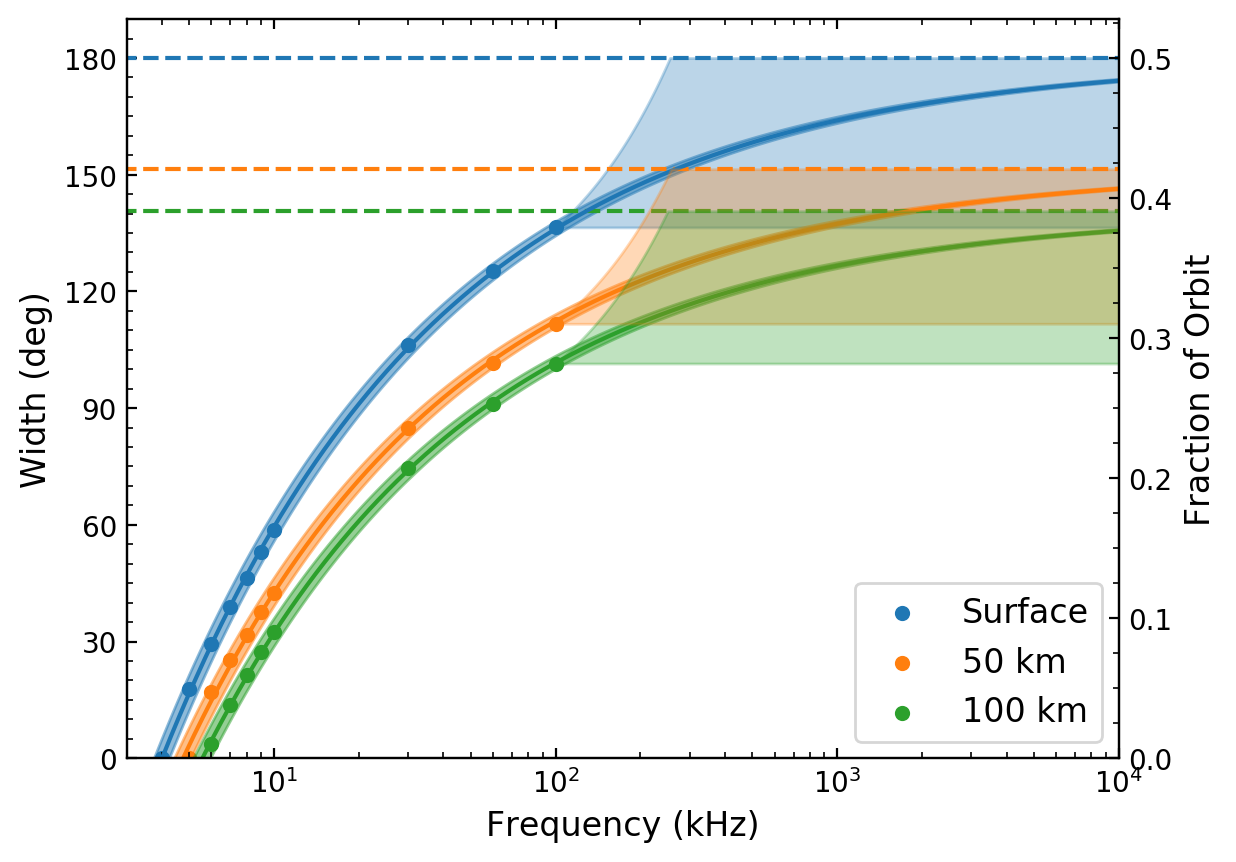}
    \caption{The width of the $\geq 80$ dB quiet region is plotted as a function of frequency for the lunar surface (blue), a circular orbit 50 km above the surface (orange), and 100 km above the surface (green). The dashed lines indicate the geometrical maximum extent of the quiet region for each of the three cases. The dots indicate FDTD simulation results. The lighter shaded regions above 100 kHz indicate the full range of possible values at frequencies above those feasibly simulated by the FDTD algorithm. The upper bounds of the shaded regions (which appear to be curved due to the log scale) are defined by extending the slope between the two highest frequency simulation points. The solid lines are produced by the best fit power law model, as described in Section \ref{extrapolation}. The darker shaded bands around the best fit lines estimate the uncertainty in the model due to numerical systematic effects as detailed in \ref{appendix}.}
    \label{fig:model}
\end{figure}

\subsection{Extrapolation to Higher Frequencies}
\label{extrapolation}

As mentioned in Section \ref{frequency}, computational resources limit the frequencies at which FDTD simulations of this system are feasible. In order to estimate the size of the quiet region at higher frequencies, we propose a model that fits the data from simulations performed at lower frequencies.

We model the width of the quiet region as a function of frequency as

\begin{equation}
\label{pwr_law}
   w = w_{max} - a\nu^{b},
\end{equation}

\noindent where $w_{max}$ is the maximum geometric width (i.e. the dashed lines in Figure \ref{fig:model}), $\nu$ is the frequency, and $a$ and $b$ are the coefficients determined by the fit. This model is motivated by the scale invariance of the system in terms of frequency. At a frequency of 10 kHz, the corresponding vacuum wavelength is 30 km. This wavelength is much smaller than the size of the Moon with a radius of 1737 km. As the frequency increases, the wavelength will become an even smaller fraction of the size of the Moon. This leads us to hypothesize a self-similar solution such that the width of the quiet region varies in proportion to the change in frequency, which is the behavior of a power law. Although the wavelength will begin to be comparable in scale to lunar features such as mountains or the depth of the regolith at higher frequencies, the skin depth at these frequencies ensures that any coupling between the lunar material and the waves will be damped out before the radiation can propagate a significant distance into the farside. Figure \ref{fig:model} shows the outcome of applying our power law model to the results from direct simulations.

\subsection{Calculating Quiet Region Width}
\label{calc_width}

The width of the quiet region depends not only on the frequency of the radiation, but also on the height above the lunar surface and the threshold chosen for the acceptable relative intensity. From the FDTD simulation results, we calculate the values of the power law coefficients $a$ and $b$ for a grid of height and intensity threshold values. 
\begin{figure*}
    \centering
    \includegraphics[scale=1]{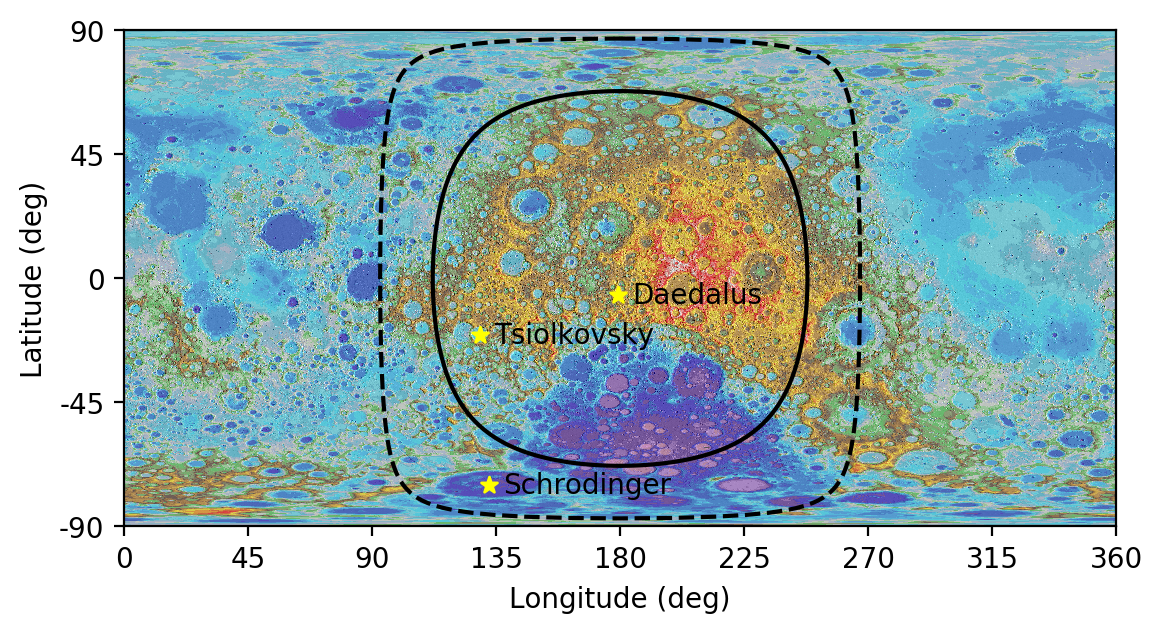}
    \caption{Elevation map from the Lunar Reconnaissance Orbiter Camera \cite[LROC;][]{Robinson:2010}. Blue and purple indicate lower elevations while yellow and red indicate higher elevations. Yellow stars signify several craters that may be of interest to low frequency radio experiments and/or lunar studies. The solid black curve denotes the $\geq$ 80 dB quiet region as determined from FDTD simulations at 100 kHz, while the dashed black curve is the $\geq$ 80 dB quiet region at 10 MHz extrapolated from the model outlined in Section \ref{extrapolation}. Note that due to libration, the quiet region will vary slightly in time, on the order of a few degrees both in Latitude and Longitude. This map was produced with the aim of identifying ideal locations for low radio frequency experiments on the lunar surface such as FARSIDE \citep{farside}.}
    \label{fig:topo_map}
\end{figure*}
We then interpolate over the grid with a two dimensional cubic spline such that the coefficient values can be calculated at any point within the height-intensity threshold parameter space. A Python package for calculating the width of the quiet region based on our FDTD simulations is made available for open-source use.\footnote{\url{https://bitbucket.org/nbassett/lunar-rfi}\label{bitbucket}} Given a frequency, height, and intensity threshold, the \texttt{calc\_width} function will calculate the width of the quiet region in degrees. In order to ensure robust results, note that the interpolation is only valid for $0 \textup{ km} < h < 150 \textup{ km}$ and $-90 < \textup{dB} < -50$.

Table \ref{tab:width_table} provides widths for a representative range of values of height and frequency. In order to estimate the uncertainty in the width, we examine the values of the power law coefficients $a$ and $b$ as a function of both height and intensity threshold. We find that the uncertainty in the values of the model coefficients are approximately $\pm 7$ for $a$ and $\pm 0.01$ for $b$ across the input parameter space. This generally corresponds to a level of uncertainty on the order of several degrees in the width of the quiet region, but decreases at higher frequencies. For further details regarding the determination of the uncertainty see \ref{appendix}.

\section{Discussion}
\label{Discussion}

21-cm observations or other low radio frequen\-cy experiments requiring a quiet environment cou\-ld be performed either directly on the lunar surface or in orbit of the Moon. In an orbiting experiment, observations would be taken while passing through the quiet region behind the farside. An orbiting experiment would be particularly conducive to a global 21-cm signal experiment, which requires only a single antenna, whereas any experiment requiring an array such as the 21-cm power spectrum or direct detection of exoplanet magnetospheres would be better suited for the lunar surface.

The map of the Moon shown in Figure \ref{fig:topo_map} reveals the large area of the farside covered by the radio quiet region. Even at frequencies as low as 100 kHz, RFI signals will be attenuated by at least 80 dB over a large portion of the farside. When the frequency is increased to 10 MHz, the quiet region covers nearly the entire farside including the South Pole Aitken Basin. This presents opportunities to deploy low frequency experiments in conjunction with lunar studies at geologically interesting locations on the farside such as the Schr\"{o}dinger crater \citep{Kramer:2013}. 

\section{Conclusions}
\label{conclusions}

We have presented numerical simulations whi\-ch model the low frequency environment behind the lunar farside. Mitigating terrestrial sources of RFI is vital for sensitive low frequency experiments such 21-cm observations or direct detection of planetary magnetospheres. Even above the Earth's ionosphere, observations would be significantly contaminated by terrestrial radio signals. The Moon, however, provides a unique environment where radio signals are heavily attenuated. This makes the lunar farside the optimal location for performing these low frequency experiments.

The simulations presented in this paper utilize a Finite Difference Time Domain (FDTD) technique for computing the time evolution of Maxwe\-ll's Equations. Average values for the permittivity and conductivity of the Moon ensure that the lunar electrical properties are integrated into the simulation. A two dimensional grid was constructed such that only a small area around the Moon was contained in the domain in order to limit the computational resources required. The consequence of this small simulation space is that the simulations only take into account the presence of the Moon and not other factors such as free space path loss, making our results a conservative estimate of the true level of attenuation of terrestrial radio sources. We note, however, that the simulations presented in this work do not take into account the tenuous lunar ionosphere or the interaction between the Moon and the Solar Wind, which could possibly affect wave propagation at kHz-scale frequencies. Our simulations also do not attempt to model the effects of AKR, which may dominate terrestrial RFI below $\sim$ 500 kHz, but are irrelevant at the higher frequencies to which our model attempts to extrapolate the size of the quiet region.

The results of our electrodynamics simulations show that as the frequency of the radiation is increased, the amount of attenuation directly behind the Moon becomes greater and the size of the region defined by a given amount of attenuation widens. This behavior is to be expected because as the frequency is increased, the waves will be diffracted by a smaller amount, which leads to greater attenuation at any given point on the farside. Although performing FDTD simulations at frequencies in the MHz range is difficult due to computational limits, results for frequencies between 5-100 kHz are able to place a lower limit on the level of attenuation. We also simulate the effect of the topography of the Moon on the intensity of the RFI by using elevation data from the LOLA instrument on the Lunar Reconnaissance Orbiter. Our results show that while the craters of the Moon do have an effect on the propagation of radio waves, the effect in the intensity of these waves on the farside is small, less than 5 dB at most.

As mentioned in Section \ref{introduction}, this work was performed to verify and extend previous studies of the lunar radio quiet region. The results from our FDTD simulations agree generally with the findings of \cite{Takahashi:2003}, which utilized a similar approach to numerically simulate the electromagnetic environment. In this work, however, we were able to directly simulate radio waves up to 100 kHz, a higher frequency than previous simulations were able to achieve. The results of this paper and \cite{Takahashi:2003} both predict a much larger amount of attenuation than that of \cite{Pluchino:2007}. On the surface of the farside, \cite{Pluchino:2007} predicted a maximum level of attenuation of $\sim 70$ dB for radiation with a frequency of 100 kHz, which is much less than the $\sim 200$ dB attenuation predicted by this work. This discrepancy can be explained, however, by the limitations of the straight edge approximation made by \cite{Pluchino:2007}. The straight edge cannot take into account the matter both in front of and behind the diffracting edge, both of which play a significant role in the attenuation of radio waves. It also assumes that the straight edge is a perfect conductor, neglecting the electromagnetic properties of the lunar material, which should affect the propagation of radio waves.

Although we limited our simulations to frequencies $\leq 100$ kHz due to computational demands, we proposed a power law model for extrapolating the width of the quiet region to higher frequencies. This model is motivated by the assumption that its frequency dependence is invariant, as described by a power law. The best fit parameters for the model were calculated for a grid of heights above the lunar surface and intensity thresholds for the quiet region. An interpolation was then performed over this grid such that the width of the quiet region can be calculated from any combination of the frequency, height, and dB threshold (within the specified bounds). In the \hyperref[bitbucket]{\texttt{lunar-rfi}} repository, we made publicly available a Python package that implements this model to calculate the size of the quiet region, with which we determine spans $172.1 \pm 0.6$ degrees at 10 MHz with a threshold of -80 dB. By directly mapping this region onto the lunar surface, we facilitate planning possible locations for future low radio frequency experiments. 

We conclude by noting that our simulations reveal that a large portion of the lunar farside is sufficiently shielded from terrestrial RFI such that sensitive low frequency observations can be performed. This quiet region provides significant opportunities for science that cannot be performed from the ground or Earth orbit. However, we note that as the circumlunar environment becomes more populated, RFI leakage from satellites and other instruments, or possibly even reflection of RFI off of lunar satellites, is increasingly likely to pollute the radio quiet environment of the farside.

\section{Acknowledgements}
\label{acknowledgements}

We would like to thank Abhirup Datta for discussions pertaining to FDTD simulations. This work is supported by the NASA Solar System Exploration Research Virtual Institute cooperative agreement number 80ARC017M0006. Reso\-urces supporting this work were provided by the NASA High-End Computing (HEC) Program through t\-he NASA Advanced Supercomputing (NAS) Division at Ames Research Center. DR was supported for part of the time by a NASA Postdoctoral Program Senior Fellowship at NASA's Ames Research Center, administered by the Universities Space Research Association under contract with the National Aeronautics and Space Administration (NASA). This work was also supported by NASA under award number NNA16BD14C for NASA Academic Mission Services.

%% The Appendices part is started with the command \appendix;
%% appendix sections are then done as normal sections
% \appendix

%% References
%%
%% Following citation commands can be used in the body text:
%% Usage of \cite is as follows:
%%   \cite{key}         ==>>  [#]
%%   \cite[chap. 2]{key} ==>> [#, chap. 2]
%%

%% References with bibTeX database:

% \bibliographystyle{elsarticle-num}
 \bibliographystyle{elsarticle-harv}

\bibliography{ref}

\onecolumn{
\appendix
\section{Model Uncertainty}
\label{appendix}

\begin{figure*}
    \centering
    \includegraphics[width=\linewidth]{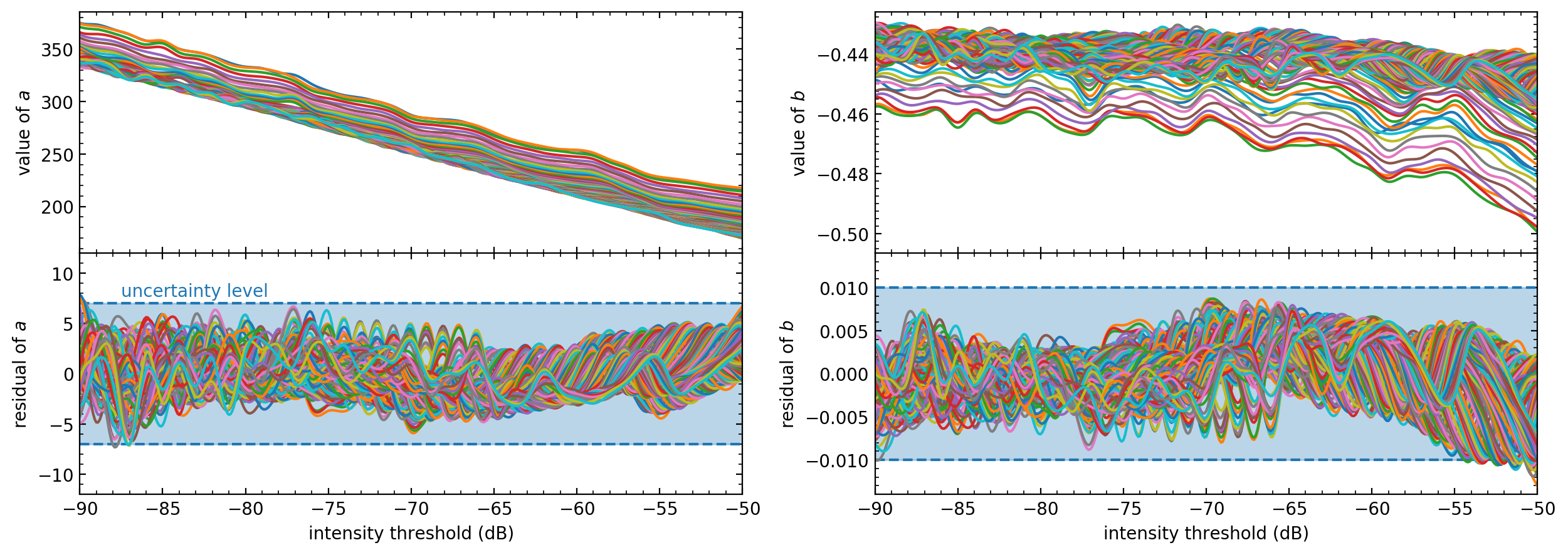}
    \caption{\textit{Top}: The value of the power law model coefficients $a$ (left) and $b$ (right) as a function of the chosen intensity threshold. Each curve corresponds to a different height above the lunar surface. \textit{Bottom}: The residuals achieved after a linear fit is removed from each of the curves in the top panels. We employ the maximum values of these residuals (approximately $\pm 7$ and $\pm 0.01$) as estimates of the uncertainties in the model parameters $a$ and $b$.}
    \label{fig:a_b_uncertainty}
\end{figure*}

As described in Section \ref{extrapolation}, the width of the quiet region depends on the frequency, height above the surface, and chosen intensity threshold. For a given height and intensity threshold, we use the power law given by Equation \ref{pwr_law} to model the frequency dependence of the quiet region. Since the coefficients $a$ and $b$ of the model are not constant, Equation \ref{pwr_law} can be re-written as
\begin{equation}
    w = w_{max}(h) - a(h, I_t)\nu^{b(h, I_t)}
\end{equation}
where $h$ is the height and $I_t$ is the intensity threshold.

The values of $a(h, I_t)$ and $b(h, I_t)$ are plotted for a range of values of $h$ and $I_t$ in the top panels of Figure \ref{fig:a_b_uncertainty}. In every case, the coefficients are approximately linear as a function of the chosen intensity threshold. We assume that any variations on top of the linear behavior are due to systematic or numerical effects. One possible systematic that may cause regular variations is the finite nature of the FDTD grid. The edge of the Moon in 2D will not be a perfect circle due to the grid spacing. These imperfections may introduce artifacts into the simulations when interacting with the propagating waves.

Under the assumption that the variations are caused by systematics or numerical effects, we fit a linear model to each curve and examine the residual (shown in the bottom panels of Figure \ref{fig:a_b_uncertainty}). We then adopt the maximum deviations in the residuals as the uncertainties in $a(h, I_t)$ and $b(h, I_t)$, which are approximately $\pm 7$ and $\pm 0.01$, respectively. When calculating the width of the quiet region, these uncertainties in the coefficients can straightforwardly be translated into uncertainties in degrees. This calculation is included as a feature in the \hyperref[bitbucket]{\texttt{lunar-rfi}} repository referenced in Section \ref{calc_width}. The small errors in this first order estimate of the uncertainty in our model lends additional reliability to the results of our simulations.
}

\end{document}